\documentclass[aps,onecolumn,superscriptaddress,nofootinbib]{revtex4-2}
\usepackage{amsmath,amssymb,graphicx,microtype,amsfonts,amstext,mathtools,physics,hyperref,bm}
\usepackage[normalem]{ulem}
\hypersetup{colorlinks=true,linkcolor=blue,citecolor=blue,urlcolor=blue}

\newcommand{\be}{\begin{equation}}
\newcommand{\ee}{\end{equation}}

\newcommand{\bse}{\begin{subequations}}
\newcommand{\ese}{\end{subequations}}
\newcommand{\bea}{\begin{eqnarray}}
\newcommand{\eea}{\end{eqnarray}}
\newcommand{\ba}{\begin{array}}
\newcommand{\ea}{\end{array}}
\newcommand{\bc}{\begin{center}}
\newcommand{\ec}{\end{center}}

\newcommand{\hairon}{\varphi_h}

\begin{document}
	
	\title{$\mathcal{H}$olographic $\mathcal{N}$aturalness and Information See-Saw Mechanism for Neutrinos}
	
	\author{Andrea Addazi}
	\email{addazi@scu.edu.cn}
	\affiliation{Center for Theoretical Physics, College of Physics Science and Technology, Sichuan University, 610065 Chengdu, China}
	\affiliation{Laboratori Nazionali di Frascati INFN, Frascati (Roma), Italy, EU}
	
	\author{Giuseppe Meluccio}
	\email{giuseppe.meluccio-ssm@unina.it}
	\affiliation{Scuola Superiore Meridionale, Largo San Marcellino 10, Napoli 80138, Italy}
	\affiliation{INFN Sezione di Napoli, Complesso Universitario di Monte Sant'Angelo, Edificio 6, Via Cintia 21, Napoli 80126, Italy}
	
	\date{\today}
	
	\begin{abstract}
		The microscopic origin of the de Sitter entropy remains a central puzzle in quantum gravity that is related to the cosmological constant problem. Within the paradigm of $\mathcal{H}$olographic $\mathcal{N}$aturalness, we propose that this entropy is carried by a vast number of light, coherent degrees of freedom---called ``hairons''---which emerge as the moduli of  gravitational instantons on orbifolds. Starting from the Euclidean de Sitter instanton ($S^4$), we construct a new class of orbifold gravitational instantons, $S^4/\mathbb{Z}_N$, where $N$ corresponds to the de Sitter entropy. We demonstrate that the dimension of the moduli space of these instantons scales linearly with $N$, and we identify these moduli with the hairon fields. A $\mathbb{Z}_N$ symmetry, derived from Wilson loops in the instanton background, ensures the distinguishability of these modes, leading to the correct entropy count. The hairons acquire a mass of the order of the Hubble scale and exhibit negligible mutual interactions, suggesting that the de Sitter vacuum is a coherent state, or Bose--Einstein condensate, of these fundamental excitations. Then, we present a novel framework which unifies neutrino mass generation with the cosmological constant through gravitational topology and holography. The small neutrino mass scale emerges naturally from first principles, without requiring new physics beyond the Standard Model and Gravity. The gravitational Chern--Simons structure and its anomaly with neutrinos force a topological Higgs mechanism, leading to neutrino condensation via $S^4/\mathbb{Z}_N$ gravitational instantons. The number of topological degrees of freedom $N \sim M_\textup{P}^2/\Lambda \sim 10^{120}$ provides both the holographic counting of the de Sitter entropy and a $1/N$ {\it information see-saw mechanism} for neutrino masses. Our framework makes the following predictions: (i) a neutrino superfluid condensation forming Cooper pairs below meV energies, as a viable candidate for cold dark matter; (ii) a possible resolution of the strong CP problem through a QCD composite axion state; (iii) time-varying neutrino masses which track the evolution of dark energy; and (iv) several distinctive signatures in astroparticle physics, ultra-high-energy cosmic rays and high magnetic field~experiments.
	\end{abstract}
	
	\maketitle
	
	\section{Introduction}
	
	The holographic nature of spacetime, first glimpsed in black hole (BH) thermodynamics, suggests that the information content of a universe dominated by a positive cosmological constant (CC) is encoded on its cosmological horizon. The de Sitter (dS) entropy, $S_\textup{dS} = A_H/(4G) = 3\pi /(G\Lambda) \equiv N\sim 10^{120}$, implies the existence of $N$ fundamental degrees of freedom, or ``qubits'', underlying the classical geometry of this spacetime. Yet, the precise microscopic identity of these degrees of freedom---the ``information atoms'' of de Sitter space---has so far remained elusive.
	
	A compelling approach, suggested by the $\mathcal{H}$olographic $\mathcal{N}$aturalness paradigm ($\mathcal{HN}$), is to look for these atoms of information not in the perturbative spectrum of quantized gravity, but in the non-perturbative structures that define the gravitational vacuum\mbox{~\cite{Addazi:2020axm,Addazi:2020wnc,Addazi:2020mnm}.} In this work, we argue that the de Sitter entropy is carried by a collection of light, weakly interacting fields, which we call ``hairons''. These hairons are not arbitrary, {ad hoc} additions; in fact, they arise naturally as the moduli, i.e., the parameters of zero-mode fluctuations, associated with a specific class of gravitational instantons. Consequently, hairons are intrinsically quantum-mechanical entities, arising solely from quantum-gravitational effects and thereby constituting the quantum ``hairs'' of spacetime.
	
	Our investigation begins with the well-known fact that the four-dimensional de Sitter space corresponds to a gravitational instanton: the Euclidean four-sphere, $S^4$, with a radius set by the cosmological constant, $R_\textup{dS} \sim 1/\sqrt{\Lambda}$~\cite{EdS}. This instanton dominates the semi-classical path integral and is thermodynamically associated with the Gibbons--Hawking temperature~\cite{EdS1,EdS2} (see also Refs.~\cite{Bousso:1998bn,Bousso:1999ms,Bousso:2000nf}).
	
	We extend this picture by constructing a new class of orbifold instantons inspired by ALE instantons: $S^4/\mathbb{Z}_N$. These are equivalent to hyperspheres with $N$ conical singularities, topologically distinct from the smooth $S^4$. A key result is that the dimension of the moduli space of these instantons, i.e., the number of independent parameters that define \mbox{them, scales} as $N$. We identify these moduli with the proposed hairon fields. A crucial element is the emergence of a $\mathbb{Z}_N$ symmetry, which we derive from the evaluation of Wilson loops wrapping around the instanton, drawing inspiration from fractional gauge instantons~\cite{FGI1,FGI2,tHooft:1981nnx,FGI3,FGI4,FGI5,FGI6,FGI7,FGI10,FGI11,FGI12,FGI13}. This symmetry guarantees that the $N$ hairons are physically distinguishable, preventing the factorial over-counting that would otherwise spoil the entropy formula.
	
	The mass of hairons is set by the Hubble scale, as $m_h \sim \sqrt{\Lambda} \sim H$ (with $H$ being the Hubble parameter), and their mutual interactions are suppressed by powers of the Planck mass. The weakness of their interactions suggests that the $N$ hairons can form a macroscopic coherent state, a Bose--Einstein condensate whose collective behavior reproduces the classical de Sitter vacuum. In this framework, the cosmological constant problem is transmuted into a question about the origin and the dynamics of these holographic degrees of freedom. Its small value is thus seen to be not fine-tuned, but simply a direct consequence of the suppression due to the enormous number $N$ of holographic degrees of freedom associated with the cosmological constant.
	
	We then discuss the possible common origin of neutrino masses and dark energy. Neutrino oscillation experiments have definitively established that neutrinos are massive, yet their sub-eV mass scale remains perplexingly small within the framework of the Standard Model. At the same time, cosmological observations depict an accelerating universe driven by dark energy, with a characteristic energy density scale $\rho_\textup{vac} \sim \text{meV}^4$. This possibly gives rise to a mysterious coincidence between the scales of neutrino masses and of dark energy: $m_\nu \sim \rho_\textup{vac}^{1/4}$.
	
	Traditional explanations for the smallness of neutrino masses invoke a see-saw mechanism, for example by introducing heavy right-handed neutrinos at scales $m_R \sim 10^{14}\divisionsymbol10^{15}$ GeV. However, this approach introduces new fundamental scales without deeper theoretical justifications, and lacks any connection to the cosmological constant problem. The holographic principle~\cite{Bousso:1998bn,Bousso:1999ms,Bousso:2000nf}, instead, suggests a radically different perspective: the cosmological constant $\Lambda$ may not be a fundamental parameter but rather an emergent quantity related to the information content of the Universe, given that the de Sitter entropy scales as $S_{\textup{dS}} \sim M_{\textup{P}}^2/\Lambda \sim N$. The relation $m_\nu \sim\rho_\text{vac}^{1/4}$ would then imply that the small neutrino masses emerge naturally from the suppression coming from the holographic de Sitter entropy $N$, effectively realizing a $1/N$ \textit{information see-saw mechanism}.
	
	In this work, we develop a comprehensive framework that unifies neutrino mass generation and dark energy through gravitational topology and holography. Our approach builds upon the following key insights.
	
	First, the topological structure of quantum gravity exhibits profound analogies with non-Abelian gauge theories~\cite{Dvali:2016uhn}. The gravitational Chern--Simons three-form $C_G$ and the topological density $ R\tilde{R}=dC_G$ play a role analogous to that of the QCD $\theta$-term. In pure gravity, the gravitational $\theta$-term would be physical, but in reality the coupling to massless neutrinos through the axial anomaly $\partial_\mu j_5^\mu \sim R\tilde{R}$ dramatically alters this picture, forcing a topological Higgs mechanism. Moreover, just like in QCD, one can envisage the existence of a composite state---analogous
	of an $\eta'$ particle---which can even be identified with the QCD axion, potentially solving the strong CP problem.
	
	Second, gravitational instantons on orbifolds $S^4/\mathbb{Z}_N$ provide the microscopic foundation for neutrino condensation. The number of topological degrees of freedom $N \sim M_{\textup{P}}^2/\Lambda$ naturally explains both the holographic counting of the de Sitter entropy and the mechanism for neutrino masses. The moduli space of these gravitational fractional instantons scales linearly with $N$, providing the hairon fields that constitute the fundamental holographic degrees of freedom. Moreover, hairons can be coupled to neutrinos, generating an effective and naturally small neutrino mass scale. Therefore, the genesis of both dark energy and neutrino mass is unified and entropically protected against large quantum radiative corrections.
	
	Third, a neutrino condensate $\langle \nu\bar{\nu} \rangle$ forms through a BCS-like mechanism mediated by attractive forces from gravitation anomalies and hairons, leading to a superfluid behavior below a critical temperature $T_c \sim \text{meV}$. This condensation spontaneously breaks global symmetries, generating both the desired mass term and a composite axion in the form of a pseudo-Nambu--Goldstone boson.
	
	Our framework exhibits several remarkable features.
	\begin{itemize}
		\item \textbf{Natural scale explanation}: The relation $m_\nu \sim \text{meV}$ emerges from first principles through a holographic scaling.
		\item \textbf{Unified mechanism}: Neutrino mass generation and dark energy share a common origin from gravitational topology and instantons.
		\item \textbf{Testable predictions}: Enhanced neutrino decays, time-varying masses, superfluid behavior, and distinctive astroparticle and laboratory signatures provide multiple experimental avenues for verification.
		\item \textbf{Theoretical consistency}: Anomaly matching conditions and topological arguments ensure the mathematical robustness of the framework.
		\item \textbf{Radical minimality}: No new heavy particles beyond the Standard Model and Gravity are required, in contrast to conventional see-saw models.
	\end{itemize}
	
	Therefore, this paper presents a systematic development of a unified framework for addressing the cosmological constant and neutrino mass problems through the principles of holography and topological quantum gravity. The argument proceeds as follows. First, the cosmological constant problem is rephrased in information-theoretic terms, establishing the holographic principle as the foundational perspective. Subsequent sections identify the microscopic degrees of freedom of the de Sitter space---the ``hairons''---by constructing a novel class of gravitational instantons on orbifolds. The physical properties and dynamics of these hairons are then analyzed, confirming their role as constituents of the de Sitter vacuum. Building upon this holographic foundation, the focus then shifts to the origin of neutrino masses. A formal analogy is drawn between the topological structures of QCD and gravity, leading to a mechanism for neutrino condensation mediated by gravitational anomalies. The theory of this condensation is developed in detail, revealing the potential for neutrino superfluidity. The phenomenological consequences of this unified picture are subsequently explored, yielding a suite of testable predictions across cosmology, astrophysics and laboratory experiments. A critical analysis of previous related models is provided, to highlight the distinct advantages of the present approach. Finally, the ``Holographic Neutrino'' paradigm is shown to synthesize all these elements, resolving outstanding problems in theoretical Physics and presenting a coherent theory whereby the cosmological constant and neutrino masses share a common origin in the topology and information content of spacetime. The paper concludes with a summary of findings and a discussion of their implications for quantum gravity and future research.
	
	\section{Holographic Information and the Cosmological Constant}
	\label{sec:holographic_cc}
	
	The holographic principle provides a profound reinterpretation of the de Sitter entropy $S_{\textup{dS}}$, suggesting that the information content of the Universe scales holographically with the horizon area:
	\begin{equation}
		\label{eq:dS_entropy}
		S_{\textup{dS}} \sim \frac{A_H}{A_{\textup{P}}} \sim \frac{M_{\textup{P}}^{2}}{\Lambda} \equiv N,
	\end{equation}
	where $A_H$ is the Hubble horizon area, $A_\textup{P}$ the Planck area, $M_\textup{P}$ the Planck mass and $\Lambda$ the CC. Here $N$ quantifies the fundamental information qubits encoded on the cosmological horizon. This relation clearly highlights the CC as an information-theoretic quantity rather than a mere vacuum energy density. The holographic relation $S_{\text{dS}} \sim M_{\text{P}}^{2}/\Lambda \equiv N$ provides a natural explanation for the smallness of the cosmological constant. In quantum gravity, the effective coupling at an energy scale $E$ is $\alpha_G(E) \sim E^{2}/M_{\text{P}}^{2}$. At the Hubble scale $E \sim \sqrt{\Lambda}$, this becomes $\alpha_G(\Lambda) \sim \Lambda/M_{\text{P}}^{2} \sim 1/N$. Thus, the observed hierarchy $\Lambda \ll M_{\text{P}}^{2}$ is not a fine-tuning but a reflection of the large number of holographic degrees of freedom associated with the de Sitter horizon. This is the core of the $\mathcal{H}$olographic $\mathcal{N}$aturalness argument: the smallness of $\Lambda$ is entropically protected by the vast information content of the vacuum.
	
	The holographic framework suggests that the vast disparity between the observed and the Planck-scale vacuum energy reflects an information-theoretic hierarchy, which we explain next.
	\begin{itemize}
		\item \textbf{Information dominance}: The vacuum information content $N \sim 10^{120}$ overwhelmingly exceeds the conventional matter-radiation entropy (the CMB entropy being $\sim 10^{88}$), positioning the CC as Universe's primary information `reservoir'.
		\item \textbf{Thermodynamic interpretation}: The square root of the CC (or the Hubble parameter $H$) emerges as the characteristic temperature of the information degrees of freedom of~vacuum:
		\begin{equation}
			T_{\textup{dS}} = \frac{H}{2\pi} = \frac{1}{2\pi}\sqrt{\frac{\Lambda}{3}},
		\end{equation}
		thus relating the vacuum energy to thermal properties of holographic data.
		\item \textbf{Inverse information scaling}: The hierarchy $\Lambda/M_{\textup{P}}^2 \sim N^{-1}$ indicates that the smallness of the CC reflects the extensive information content encoded holographically in the de Sitter space.
	\end{itemize}
	
	In particular, vacuum transitions between different cosmological constant states would represent information-theoretic `catastrophes'. Consider for example the transition
	\begin{equation}
		\label{eq:vacuum_transition}
		|\Lambda\rangle \rightarrow |\Lambda_{\text{UV}}\rangle \quad \Leftrightarrow \quad |N\rangle \rightarrow |1\rangle,
	\end{equation}
	where $\Lambda_{\text{UV}} \sim M_{\textup{P}}^2$ corresponds to a near-unit entropy. This process would annihilate $N-1$ information qubits, but is subject to an exponential suppression:
	\begin{equation}
		\label{eq:entropic_suppression}
		\langle \Lambda|\Lambda_{\text{UV}}\rangle \sim e^{-S_{\textup{dS}}/2} \sim e^{-N/2} \sim \exp\left(-10^{120}/2\right).
	\end{equation}
	
	This entropic protection finds an instantonic realization through the gravitational Euclidean action:
	\begin{equation}
		\label{eq:instanton_suppression}
		e^{-S_E} \sim e^{-1/\alpha_G(\Lambda)}, \quad \text{with} \quad \alpha_G(\Lambda) = \frac{\Lambda}{M_{\textup{P}}^{2}} \sim N^{-1},
	\end{equation}
	where $\alpha_G(E^2)$ is the gravitational coupling at an energy scale $E$. Remarkably, the value $\alpha_G(\Lambda)$ exhibits a distinguished information-theoretic behavior: its inverse counts the holographic degrees of freedom of the dS vacuum, intimately linking the strength of the gravitational interaction to the vacuum information content.
	
	\section{Hairons, Orbifold Instantons and de Sitter Entropy}
	\label{sec:hairons}
	
	One intriguing possibility is that hairons are related to the moduli of non-perturbative effects, specifically gravitational instantons on orbifold spaces. We propose that a $\mathbb{Z}_N$ symmetry, where $N$ is connected to the de Sitter entropy, can be derived from the evaluation of Wilson loops wrapping around  gravitational instantons.
	
	It is well established that the dS spacetime corresponds to a gravitational instanton solution, which is simply the Euclidean dS with the topology of a four-sphere, $S^4$, with radius $R \sim 1/\sqrt{\Lambda}$~\cite{EdS}. This solution is obtained via a Wick rotation of the time coordinate, $t \rightarrow i\tau$, where the period of this coordinate is identified with the inverse Gibbons--Hawking temperature, $\beta = 1/T_\textup{dS} \sim 1/\sqrt{\Lambda}$~\cite{EdS1,EdS2} (see also Refs.~\cite{Bousso:1998bn,Bousso:1999ms,Bousso:2000nf}).
	However, such dS spacetime has a large entropy scaling which is typically not understood in terms of information of qubit content. In the following we will discuss how a nearly dS space can emerge from considering a large number of Euclidean conic singularities or conifolds, which are topologically equivalent to the $S_{4}/\mathbb{Z}_{N}$ orbifold. We will start with a discussion about fractional gauge instantons and the emergence of $\mathbb{Z}_{N}$ symmetries.
	
	\subsection{Wilson Loops, Instantons and $\mathbb{Z}_N$ Symmetry}
	
	In gauge theories, the expectation value of a Wilson loop in the presence of an instanton is given by the path integral
	\begin{equation}
		\label{eq:wilson_loop}
		\langle W[\gamma] \rangle = \frac{\int \mathcal{D}A \, W[\gamma] \, e^{-S[A]} \delta(F[A] - \tilde{F}[A])}{\int \mathcal{D}A \, e^{-S[A]} \delta(F[A] - \tilde{F}[A])}.
	\end{equation}
	Here, $S[A]$ is the Yang--Mills action and the delta function $\delta(F[A] - \tilde{F}[A])$ enforces the self-duality condition $F_{\mu\nu} = \tilde{F}_{\mu\nu}$ for the instanton. For an instanton of topological charge $Q = K/N$, the classical action is quantized and proportional to the charge:
	\begin{equation}
		\label{eq:instanton_action}
		S[A] = \frac{8\pi^2}{g^2} \frac{K}{N} \sim \frac{1}{\alpha_\text{YM}}\frac{K}{N}.
	\end{equation}
	For large Wilson loops, this leads to a non-trivial phase:
	\begin{equation}
		\label{eq:wilson_phase}
		\langle W[\gamma] \rangle \sim \exp\left( -2\pi i \frac{K}{N} \right),
	\end{equation}
	where $N$ is the rank of the gauge group. This phase factor $\exp(-2\pi i K/N)$ stems from the non-trivial holonomy of the gauge field around the loop due to the instanton's topological charge. Such configurations are known as fractional gauge instantons~\cite{FGI1,FGI2,tHooft:1981nnx,FGI3,FGI4,FGI5,FGI6,FGI7,FGI10,FGI11,FGI12,FGI13}, typically arising on manifolds with twisted boundary conditions. The residual center symmetry of $SU(N)$ in this context is $\mathbb{Z}_N$.
	
	\subsection{Orbifold Gravitational Instantons and ALE Spaces}
	
	The gauge theory picture suggests the existence of a new class of gravitational instantons atop the standard Euclidean dS one ($S^4$). These are the orbifolds of the hypersphere, $O_{4,N} = S^4 / \mathbb{Z}_N$. The space $O_{4,N}$ is a hypersphere with $N$ conical singularities, which correspond topologically to a set of two-surfaces $\Sigma$.
	
	This new class of instantons is related to the well-known Asymptotically Locally Euclidean (ALE) instantons~\cite{ALE1,ALE2,ALE3,ALE4,ALE5}. The connection emerges through a local analysis near the orbifold singularities. While $S^4/\mathbb{Z}_N$ is a compact orbifold, the geometry near each singularity is locally modeled by $\mathbb{R}^4/\mathbb{Z}_N $, which can be complexified to $\mathbb{C}^2/\mathbb{Z}_N$. Instantons on $S^4/\mathbb{Z}_N$, when lifted to the covering space $S^4$ and projected to $\mathbb{R}^4$ via stereographic projection, yield $\mathbb{Z}_N$-symmetric instanton configurations on $\mathbb{R}^4$. The local behavior near a fixed point defines an orbifold instanton on $\mathbb{R}^4/\mathbb{Z}_N$, which represents the singular limit of a smooth, hyper-K\"ahler ALE instanton upon resolution of the singularity. Thus, the global $S^4/\mathbb{Z}_N$ instanton provides a compact framework encapsulating the local data of $A_{N-1}$-type ALE instantons at its singular points---Gravitational instantons asymptotic to $\mathbb{C}^2/\mathbb{Z}_N$---with the fractional instanton numbers accounted for by the Kronheimer--Nakajima construction upon desingularization~\cite{ALE1,ALE4,ALE5}.
	
	\subsection{Moduli Space and Conical Singularities}
	
	A key difference between the $S^4$ and $S^4/\mathbb{Z}_N$ instantons lies in their moduli spaces. The moduli space for $S^4$ instantons is trivial, corresponding to the instanton's center and having dimension $\text{dim}(S^4)=4$, which is the dimension of the coset $SO(5)/SO(4)$ (with $SO(5)$ being the isometry group of $S^4$). This does not match the moduli count for $S^4/\mathbb{Z}_N$, which instead scales as $N$. More precisely, $S^4/\mathbb{Z}_N$ has a moduli space dimension equivalent to that of $\mathbb{Z}_N$ ALE instantons. For an instanton winding number $K=1$, this dimension is $\dim \mathcal{M}(S^4, N) \sim 3N$, counting the positions of the $N$ fixed orbifold singularities. This leads to a corresponding number of Nambu--Goldstone bosons associated with the spontaneous breaking of isometries, which scales as $N$ too.
	
	The geometry of $S^4/\mathbb{Z}_N$ is that of $N$ conical singularities characterized by a deficit angle $\Theta$ and an opening angle $\Psi$:
	\begin{equation}
		\label{eq:deficit_angle}
		\Theta = 2\pi( 1 - \Psi), \quad \text{with} \quad \Psi = 8\pi G\mu \sim \frac{1}{N}.
	\end{equation}
	Here, $G$ is the gravitational constant and $\mu$ the Euclidean string tension that sustains the conical geometry. In the dS case, this tension is proportional to the CC, $\mu \propto \Lambda$. In the limit $N \rightarrow \infty$, the standard Euclidean dS geometry is recovered, as $\Theta_\textup{dS}=2\pi$ reflects the periodicity of its Euclidean time coordinate. Equation \eqref{eq:deficit_angle} implies an intriguing relation:
	\begin{equation}
		\label{eq:uncertainty}
		\Psi S_\textup{dS} \sim 1,
	\end{equation}
	which can be interpreted as an uncertainty relation between the opening angle and the dS entropy or, equivalently, the number of fundamental qubits of information in vacuum.
	
	\subsection{A Geometric Construction of Gravitational Orbifold Instantons}
	In this subsection we now discuss two possible mathematical constructions for the $S^4/\mathbb{Z}_N$ gravitational instantons.
	
	To begin with, consider the round 4-sphere of radius $R=1$ with metric
	\begin{equation}
		\label{st11}
		ds^2_{S^4} = d\chi^2 + \sin^2\chi\left[(d\psi + \cos\theta\, d\phi)^2 + d\theta^2 + \sin^2\theta\, d\phi^2\right],
	\end{equation}
	with $0 \le \chi \le \pi$, $0 \le \theta \le \pi$, $0 \le \phi < 2\pi$ and $0 \le \psi < 4\pi$. Such a metric can also be written in a conformal form as
	\begin{equation}
		\label{standard}
		ds^2_{S^4} = \frac{4R^4}{(R^2 + |z|^2)^2} \left( |dz_1|^2 + |dz_2|^2 \right),
	\end{equation}
	with $R=1$ and $|z|^2 = \sin^2\chi$, where
	\begin{equation}
		\label{zou}
		z_1 = \sin\chi \cos(\theta/2) \, e^{i(\psi+\phi)/2}, \qquad 
		z_2 = \sin\chi \sin(\theta/2) \, e^{i(\psi-\phi)/2},
	\end{equation}
	with $|z_1|^2 + |z_2|^2 = \sin^2\chi$. These coordinates contain the standard Hopf coordinates for $S^3$, which read
	\begin{equation}
		\label{zuzd}
		\tilde{z}_1 = \cos(\theta/2) \, e^{i(\psi+\phi)/2}, \qquad 
		\tilde{z}_2 = \sin(\theta/2) \, e^{i(\psi-\phi)/2}.
	\end{equation}
	In fact, such coordinates correspond to the Hopf fibration, which is $S^1 \hookrightarrow S^3 \xrightarrow{p} S^2$, where the total space is $S^3 \subset \mathbb{C}^2$ with $|z_1|^2 + |z_2|^2 = 1$. As is known, contrary to $S^3$, there is not any defined Hopf fibration $S^1 \to S^4 \to M^3$ with $M^3$ being a smooth manifold. However, we can still parametrize $S^4$ using coordinates that resemble the Hopf fibration on each $S^3$ slice.
	
	The $S^4/\mathbb{Z}_N$ instantons can be constructed on the basis of the metrics \eqref{standard}. For the $\mathbb{Z}_N$ action, defined by the symmetry
	\begin{equation}
		\psi \mapsto \psi + \frac{4\pi}{N}, \qquad \text{with } (\chi,\theta,\phi) \text{ fixed},
	\end{equation}
	the corresponding action in complex coordinates $(z_1, z_2)$ is given by
	\begin{equation}
		\label{zkllz}
		(z_1, z_2) \mapsto (e^{2\pi i/N} z_1, e^{2\pi i/N} z_2).
	\end{equation}
	This gives two fixed points at the pole $z_1 = z_2 = 0$ and at its antipodal point (i.e., the south pole after compactification).
	
	Let us then introduce the quotient coordinate $\tilde{\psi} \equiv N\psi$ with period $4\pi$. The resulting orbifold metric is
	\begin{equation}
		\label{orb}
		ds^2_{\text{orb}} = d\chi^2 + \frac{\sin^2\chi}{N^2}(d\tilde{\psi} + N\cos\theta\, d\phi)^2 + \sin^2\chi\,(d\theta^2 + \sin^2\theta\, d\phi^2).
	\end{equation}
	For $\chi = 0$, the prefactor $\sin^2\chi$ in front of $(d\theta^2 + \sin^2\theta\, d\phi^2)$ vanishes, implying that $S^2$ collapses to a point. Therefore, the fixed set is the entire $S^2$ at $\chi = 0,\pi$, but this $S^2$ is of zero measure in the metric---it is a single `point' for the metric. This is thus a codimension-2 singular locus that is collapsed to codimension-4 in a geometric way.
	
	As for the singularity structure, near the poles $\chi \to 0$ and $\chi \to \pi$ we find that
	\[
	ds^2_{\text{orb}} \approx d\chi^2 + \frac{\chi^2}{N^2}(d\tilde{\psi} + N\cos\theta\, d\phi)^2 + \chi^2(d\theta^2 + \sin^2\theta\, d\phi^2),
	\]
	revealing two singular $S^2$ at $\chi=0$ and $\chi=\pi$, each with a conical deficit of $2\pi(1-1/N)$ in the normal plane. The normal bundle to each singular $S^2$ has Chern class $N$.
	
	We now move on to an alternative geometric construction. The ALE gluing construction starts from the observation that $S^4$ minus two antipodal points is conformally equivalent to $\mathbb{R}^4 \setminus \{0\} \cong \mathbb{R}^+ \times S^3$. By considering an appropriate $\mathbb{Z}_N$-action on $\mathbb{C}^2$, we obtain a picture where singularities are isolated points that can be resolved by hyper-k\"{a}hler ALE spaces. The main ingredients of our derivation are the following.
	\begin{itemize}
		\item {\it Einstein metric:} As a quotient of an Einstein space, it satisfies $R_{\mu\nu} = (3/R^2)g_{\mu\nu}$ away from singularities.
		\item {\it Self-duality:} The Weyl tensor vanishes identically (conformally flat quotient).
		\item {\it Global structure:} The singular locus consists of two antipodal 2-spheres.
		\item {\it Boundary at infinity:} Non-existent, i.e.,\ a compact space.
	\end{itemize}
	This construction emphasizes that $S^4/\mathbb{Z}_N$ is naturally an \textit{orbifold Einstein space} with a specific singular geometry, which can be built in four steps as we show next.
	
	{\it --- Step 1}: stereographic coordinates and $\mathbb{Z}_N$-action. Let $S^4 = \mathbb{R}^4 \cup \{\infty\}$ via stereographic projection. Then, in complex coordinates $(z_1, z_2) \in \mathbb{C}^2 \cong \mathbb{R}^4$, define
	\[
	(z_1, z_2) \mapsto (\omega z_1, \omega^{-1} z_2), \qquad \omega = e^{2\pi i/N}.
	\]
	Such an action fixes only $(0,0)$ in $\mathbb{R}^4$ and the point at infinity in $S^4$.
	
	{\it --- Step 2}: excising singular neighborhoods. We then remove the geodesic balls $B_\epsilon(p_+)$ and $B_\epsilon(p_-)$ around the poles (with $\epsilon>0$). Each boundary is thus the lens space \linebreak  $\partial B_\epsilon \cong S^3/\mathbb{Z}_N = L(N,1)$.
	
	{\it --- Step 3}: ALE replacements. Given two copies of the $A_{N-1}$ ALE space, the minimal resolution of $\mathbb{C}^2/\mathbb{Z}_N$ with hyper-k\"{a}hler metric (according to the Gibbons--Hawking ansatz)~is
	\begin{align*}
		ds^2_{\text{ALE}} &= V(\mathbf{x})^{-1}(d\tau + \boldsymbol{\omega}\cdot d\mathbf{x})^2 + V(\mathbf{x})d\mathbf{x}^2, \\
		V(\mathbf{x}) &= \sum_{i=1}^{N} \frac{1}{|\mathbf{x} - \mathbf{a}_i|}, \quad \nabla \times \boldsymbol{\omega} = \nabla V.
	\end{align*}
	Each ALE can then be truncated at a large radius $r=R_0$, creating a boundary $\partial_{\text{ALE}} \cong L(N,1)$.
	
	{\it --- Step 4}: gluing conditions. Finally, we can glue along boundaries with the matching conditions
	\begin{enumerate}
		\item {\it Induced metric match:} $h_{ij}^+ = h_{ij}^-$ on $L(N,1)$;
		\item {\it Extrinsic curvature match:} $K_{ij}^+ = K_{ij}^-$ for a smooth Einstein metric.
	\end{enumerate}
	These conditions impose constraints on the ALE moduli $\mathbf{a}_i$ and the gluing parameter $\epsilon$.
	
	Furthermore, the moduli space dimension arises from the counting
	\begin{equation}
		\label{MNN}
		\dim \mathcal{M}_N = 2 \times \underbrace{4(N-1)}_{\text{ALE moduli}} - \underbrace{6}_{\text{gluing constraints}} - \underbrace{6}_{\text{gauge}} + \delta = 8N - 20 + \delta,
	\end{equation}
	where $\delta$ includes topological corrections. For large $N$, we see that $\dim \mathcal{M}_N \sim 8N$.
	
	The key properties of these solutions are the following.
	\begin{itemize}
		\item {\it Smoothness:} Achieved by resolution of singularities.
		\item {\it Self-duality:} Preserved by an appropriate choice of ALE moduli.
		\item {\it Not Einstein (generically):} The Ricci tensor has a delta-function support on gluing surfaces unless fine-tuned.
		\item {\it Explicit parameters:} The moduli correspond to positions of ALE centers and relative orientations.
	\end{itemize}
	This second construction thus emphasizes the smooth resolution and the parameter space of near-instanton configurations.
	
	Both constructions presented in this subsection yield the same underlying topological orbifold $S^4/\mathbb{Z}_N$. The difference lies in the \textit{nature of the singularities}.
	\begin{itemize}
		\item {\it Hopf construction:} The singularities are of \textit{codimension-2} (two $S^2$).
		\item {\it ALE construction:} The singularities are of \textit{codimension-4} (two points) after a choice of coordinates.
	\end{itemize}
	These are related by a change in the $\mathbb{Z}_N$-action:
	\begin{itemize}
		\item The Hopf action $(z_1, z_2) \mapsto (\omega z_1, \omega z_2)$ gives codimension-2 singularities.
		\item The diagonal action $(z_1, z_2) \mapsto (\omega z_1, \omega^{-1} z_2)$ gives codimension-4 singularities.
	\end{itemize}
	The Hopf quotient metric can be obtained from the ALE-glued metric in the so-called \textit{blow-down limit} where ALE scales vanish. Conversely, the ALE-glued metric represents a \textit{smooth deformation} of the singular orbifold. The two geometric constructions of $S^{4}/\mathbb{Z}_{N}$ instantons differ in the chosen $\mathbb{Z}_{N}$ action on the complex coordinates $(z_{1}, z_{2})$. The Hopf action $(z_{1}, z_{2}) \mapsto (\omega z_{1}, \omega z_{2})$ yields singularities of codimension-2 (two $S^{2}$), while the diagonal action $(z_{1}, z_{2}) \mapsto (\omega z_{1}, \omega^{-1} z_{2})$ gives isolated fixed points (codimension-4). These are merely different coordinate representations of the same underlying orbifold. The ALE construction resolves the singularities into smooth hyper-K\"ahler geometries, providing a physically regular instanton solution. The essential topological property---the existence of $N$ conical defects---is independent of the coordinate description.

	\subsection{Metric, Periodicity and Entropy from Cones}
	
	The metric for the $(3+1)$-dimensional Euclidean de Sitter space is
	\begin{equation}
		\label{eq:eds_metric}
		ds_{E}^2 = \left(1-\frac{r^{2}}{l^{2}}\right)dt_{E}^{2} + \left(1-\frac{r^{2}}{l^{2}}\right)^{-1}dr^{2} + r^{2}d\Omega_{2}^2,
	\end{equation}
	with $t_{E} \rightarrow t_{E} + \beta$, $\beta = 2\pi l = 1/T_\textup{dS}$ and $l \sim 1/\sqrt{\Lambda}$. A coordinate transformation $r = l \cos \rho$ rewrites this as the standard four-sphere metric:
	\begin{equation}
		\label{eq:sphere_metric}
		ds_{E}^{2} = l^2 \left( \sin^{2}\rho \, dt_{E}^{2} + d\rho^2 + \cos^{2}\rho \, d\Omega_{2}^2 \right).
	\end{equation}
	To introduce a conical singularity at $r=0$ while keeping the boundary $r=l$ regular, the periodicity of $t_E$ must be altered to
	\begin{equation}
		\label{eq:zn_periodicity}
		t_{E} \rightarrow t_{E} + \frac{2\pi l}{N},
	\end{equation}
	which corresponds to the $\mathbb{Z}_{N}$ symmetry shift. In the limit $N \rightarrow \infty$, the periodicity vanishes. The replacement $\beta \rightarrow \beta/N$ in the partition function $Z = e^{-\beta F}$ (where $F=E - S/\beta$ is the free energy) implies that the partition function of each cone is a $1/N$ power of the total dS partition function:
	\begin{equation}
		\label{eq:partition_relation}
		Z_\textup{dS} \simeq (Z_\text{cone})^{N}, \quad \text{thus} \quad S_\text{EdS} \simeq S_{S^{4}/Z_{N}} \simeq N S_\text{cone},
	\end{equation}
	with $S_\text{cone} \sim 1$. This result indicates that the dS spacetime and its entropy emerge from the superposition of a large number $N$ of gravitational conical instantons. These considerations are substantiated in the lower-dimensional $\text{dS}_3$ space, where the correspondence between conical instantons and dS entropy has been proven through exact computations and confirmed via the dual $\text{CFT}_2$~\cite{Banados:1998tb}. Furthermore, this picture agrees with wave function formulations that promote the Wald entropy and the opening angle to conjugate quantum operators~\cite{Brustein:2012sa}.
	
	Following this logic, the number of hairons scales with the number of singular points on the orbifold, establishing the correspondence
	\begin{equation} 
		N\ \textbf{hairons} \leftrightarrow N\ \textbf{conic singularities} \leftrightarrow S^{4}/\mathbb{Z}_{N}.
	\end{equation}
	
	The moduli fields emerging from this analysis have clear physical interpretations, as follows.
	\begin{itemize}
		\item Each modulus corresponds to a zero mode of the linearized gravitational equations around the instanton background.
		\item These modes are Nambu--Goldstone bosons associated with the spontaneous breaking of symmetries by the instanton solution.
		\item The $\mathbb{Z}_N$ distinguishability of the moduli ensures the correct entropy counting without factorial suppressions.
		\item The mass scale $m_h \sim \sqrt{\Lambda}$ emerges from non-perturbative effects that lift the would-be flat directions of the moduli space.
	\end{itemize}

	\subsection{Qubits as Hairons}
	
	A key insight is that these modes act as natural qubits for storing information about the region of the dS spacetime which is inaccessible to a given observer~\cite{Dvali:2015onv}. The information storage capacity is quantified by the dimensionless measure
	\begin{equation}
		\label{eq:information_capacity}
		\mathcal{I} \equiv \frac{\hbar}{l\epsilon},
	\end{equation}
	where $\epsilon$ is the energy gap of the hairons and, only in this subsection, factors $\hbar$ are restored.
	
	For a massless gauge theory like linearized gravity, quantum effects generate a gap:
	\begin{equation}
		\label{eq:stuckelberg_gap}
		\epsilon= \alpha(l)\frac{\hbar}{l},
	\end{equation}
	where $\alpha(l)$ is the quantum coupling at the length scale $l$. This leads to the universal information measure
	\begin{equation}
		\label{eq:universal_information}
		\mathcal{I} = \frac{1}{\alpha(l)}.
	\end{equation}
	In the case of the gravitational interaction, the coupling scales as $\alpha_G(l) = G/l^2 \sim L_\textup{P}^2/l^2$ with $L_\textup{P}$ being the Planck length, thus yielding
	\begin{equation}
		\label{eq:gravitational_information}
		\mathcal{I}_\textup{dS} = \frac{l^2}{L_P^2} \sim S_{\textup{dS}}.
	\end{equation}
	Therefore, the de Sitter entropy precisely equals the information capacity of the hairon qubits associated with the horizon, which explains the emergence of the Entropy-Area law from the perspective of information theory.
	
	The quantity $\mathcal{I}$ defined in Equation\ \eqref{eq:information_capacity} provides a direct measure of the information storage capacity of hairon modes. To understand this connection, consider that storing a single bit of information in a generic quantum system of size $l$ typically requires an energy of order $\hbar/l$, which represents the natural level spacing. The ratio $\mathcal{I} = \hbar/(l\epsilon)$ then counts how many hairon qubits can be excited within this fundamental energy budget.
	
	The entropy contribution from the hairon sector can be computed by counting the number of accessible states. If we restrict ourselves to the first excited level for each qubit, a state with $n$ excited qubits has energy $E = n\epsilon$. Hence, imposing the constraint $E < \hbar/l$ gives the maximum number of excitable qubits: $n_{\text{max}} = \mathcal{I}$. The number of such states is approximately $2^{\mathcal{I}}$, which yields an entropy $S \approx \mathcal{I}$.
	
	A more precise combinatorial counting accounts for all the states with total occupation number $n \leq \mathcal{I}$. For $N$ distinguishable  modes, the number of states with exactly $n$ excited qubits is given by the binomial coefficient $\binom{N+n-1}{n}$, which counts the number of ways to distribute $n$ excitations among $N$ distinguishable modes. Summing over all possible occupation numbers up to $\mathcal{I}$ gives the total number of accessible states:
	\begin{equation}
		\label{eq:state_counting}
		e^{S} = \sum_{n=0}^{\mathcal{I}} \binom{N+n-1}{n} = \frac{\Gamma(1+N+\mathcal{I})}{\Gamma(1+\mathcal{I})\Gamma(1+N)},
	\end{equation}
	where the second equality follows from combinatorial identities involving the Gamma function $\Gamma$.

	\section{Mass and Dynamics of Hairons}
	\label{sec:hairon_dynamics}
	
	The physical properties of the proposed hairon fields, identified with the moduli of the $S^4/\mathbb{Z}_N$ orbifold instantons, can be determined by analyzing the fluctuations around this gravitational background. The spectrum of metric perturbations is governed by the Lichnerowicz operator, while the dynamics and interactions of the moduli fields themselves are derived from the geometry of the instanton's moduli space.
	
	\subsection{Hairon Dynamics and Cosmological Constant}
	
	One simple mechanism to generate a mass for the hairons is through the non-minimal coupling
	\begin{equation}
		\label{coupl}
		-\int d^{4}x\sqrt{-g} \frac{\zeta}{2} \varphi_{i}\varphi^{i} R \rightarrow - \int d^{4}x\sqrt{-g} \frac{\zeta}{2} \varphi_{i}\varphi^{i} (\bar{R}+\delta R),
	\end{equation}
	where $R=\bar{R}+\delta R$, $\bar{R}=4\Lambda=12 H^2$ and $\zeta$ is a coupling constant. This translates into an effective mass term for the hairons with
	\begin{equation}
		\label{ffaa}
		m_{h}= 2\sqrt{\zeta \Lambda} \, . 
	\end{equation}
	For this mechanism to actually work, the Universe must have begun with a small but nonzero CC. A theoretical framework known as Pre-Geometric Gravity offers a way to achieve this. In this model, in fact, a ``gravitational Higgs'' field spontaneously breaks the fundamental symmetry of spacetime from a larger gauge group down to the familiar Lorentz symmetry~\cite{wilczek:gauge,Addazi:2024rzo,Addazi:2025vbw,meluccio-d,addazi:2512.20681}. One of the key outcomes of this process is the generation of a small CC. The smallness of this constant is a consequence of a natural see-saw mechanism, as its value is suppressed by the very large vacuum expectation value of the gravitational Higgs field~\cite{Addazi:2024rzo}.
	
	In addition, while the hairon moduli are classically zero modes, non-perturbative effects can generate a small mass too. This is analogous to axion mass generation in gauge theories due to instanton configurations.
	
	In the gravitational context, we can consider the interaction portal of hairons with the topological term $R\tilde{R}$. The generation of a periodic potential is suggested by the existence of an approximate $\mathbb{Z}_{N}$ symmetry, which can be related to a multi-vacua structure. This can generate a potential on the moduli space. In particular, hairons can also be thought of as complex fields with an imaginary part related to the deficit angle of conic singularities. The semi-classical contribution to the effective action, in fact, takes the form~\cite{Chen:2021jcb}
	\begin{equation}
		\label{EoMM}
		\int d^4x \sqrt{-g} \, \left[ - M_{\textup{P}}^{4}  \, e^{-2S_0} \cos\left(N \frac{b}{M_\textup{P}} + \theta\right) + \cdots \right],
	\end{equation}
	where $b\equiv\text{Im} (\sqrt{\varphi_{I}\varphi^{I}})$ is invariant under the shift symmetry 
	\begin{equation}
		\label{shift}
		b\rightarrow b +\frac{2\pi}{N}\, . 
	\end{equation}
	These contributions will be relevant only if the non-perturbative corrections match the $\Lambda$~scale:
	\begin{equation}
		\label{scal}
		\Lambda= M_\textup{P}^2 e^{-2S_{0}},\,\,\, S_{0} \simeq \frac{1}{2}\log N \simeq 120\, .
	\end{equation}
	Let us also note that the $\theta$ angle appearing in Equation\ \eqref{EoMM} is, in this geometric interpretation, related to the opening angle of the conic instantons. 
	
	The hairon dynamics are governed by the geometry of the moduli space $\mathcal{M}_{S^4/\mathbb{Z}_N}$. The low-energy effective action for the hairons takes the form of a typical non-linear sigma~model:
	\begin{equation}
		S_{\text{hairon}} = \frac{1}{2} \int d^4x \sqrt{g} \, G_{ij}(\hairon) \nabla_\mu \hairon^i \nabla^\mu \hairon^j + \cdots,
	\end{equation}
	where $G_{ij}(\hairon)$ is the metric on moduli space. The interactions between hairons are suppressed by the Planck scale. Expanding around a reference point in the moduli space~\cite{Ketov}, the leading interactions are 
	\begin{equation}
		\mathcal{L}_{\text{int}} = \frac{1}{M_\textup{P}^2} \Gamma_{ijk}^l \nabla_\mu \hairon^i \nabla^\mu \hairon^j \hairon^k \end{equation}
	$$+ \frac{1}{M_\textup{P}^4} R_{ijkl} \nabla_\mu \hairon^i \nabla^\mu \hairon^j \nabla_\nu \hairon^k \nabla^\nu \hairon^l + \cdots,$$
	where $\Gamma_{ijk}^l$ \textls[-15]{are the Christoffel symbols and $R_{ijkl}$ the Riemann tensor of the moduli space~metric.}
	
	The effective coupling for $2 \to 2$ scattering at energies $E \sim H$ is
	\begin{equation}
		\lambda_{\text{eff}} \sim \left( \frac{H}{M_\textup{P}} \right)^4 \sim \frac{1}{N^2} \ll 1.
	\end{equation}
	The extreme weakness of interactions validates the use of a dilute gas approximation for the hairon fields.
	
	\subsection{Coherent State Formation and Bose--Einstein Condensate}
	The introduction of an $S_{4}/\mathbb{Z}_{N}$ instanton is related to a large number of twisted moduli fields, the hairons, which have different fractional charges with respect to $\mathbb{Z}_N$, transforming under the symmetry generator as
	\begin{equation}
		\label{fractttion}
		g_{k}: \varphi_{h}^{k}\rightarrow e^{2\pi i k/N}\varphi_{h}^{k}\,,
	\end{equation}
	with fractional charges $q_{k}\equiv k/N$. This introduces a distinguishability in the hairon statistics which is also essential for reproducing the holographic information scaling. In this sense, an ensamble of cold hairons would appear as a cold gas rather than a condensate, with an average temperature of $T_\textup{dS}\sim \sqrt{\Lambda}\sim M_\textup{P}/\sqrt{N}$. Nevertheless, an expectation value like
	\begin{equation}
		\label{hairoon}
		\langle \hat{\hairon}^i(x) \rangle = \varphi_0^i,
	\end{equation}
	which would blow up the singular points introduced by the orbifold, would not be compatible with the $\mathbb{Z}_N$ symmetry. A vacuum expectation value that preserves the $\mathbb{Z}_N$ symmetry is instead
	\begin{equation}
		\langle |\hat{\hairon}|^2 \rangle = |\varphi_0|^2,
	\end{equation}
	which requires the existence of negatively charged twisted moduli or negative-charge hairons. In fact, both charge orientations or signs are possible for $S_4/\mathbb{Z}_N$ and therefore we will consider a superposition state of the two corresponding instantons:
	\begin{equation}
		\label{superr}
		S_{4}^{+}/\mathbb{Z}_{N}+S_{4}^{-}/\mathbb{Z}_{N}\,,
	\end{equation}
	where the superscripts ``$\pm$'' indicate the sign of the fractional charge. In this scenario, hairons--antihairons or positive--negative-charge hairons can form neutral pairs, as a sort of Cooper pairs, as their annihilation channels into other particles is highly suppressed---they are very weakly coupled to the fields of the Standard Model. The recombination of hairons with the same modulus charge $|q_{k}|$ renders all the pairs indistinguishable bosons. Their weak interactions thus allow the $N$ hairon fields to form a macroscopic Bose--Einstein condensate. The excitations of this condensate follow a Bose--Einstein distribution for modes with momentum $k > 0$:
	\begin{equation}
		n_k \sim \frac{1}{e^{\beta \omega_k} - 1}, \qquad \text{with } \beta = \frac{1}{T_\textup{dS}} = \frac{2\pi}{H}.
	\end{equation}
	The key to the phenomenon of condensation is the zero-mode sector, where the energy cost $\omega_0$ is negligible compared to the temperature $T_\textup{dS} = H/2\pi$. In this limit ($\beta \omega_0 \to 0$), the occupation number becomes
	\begin{equation}
		n_0 \sim \frac{1}{\beta \omega_0} = \frac{T_\textup{dS}}{\omega_0} \gg 1.
	\end{equation}
	With a large number of fields $N$, the macroscopic occupation number $n_0$ can be of order $N$, indicating Bose--Einstein condensation indeed. In the case of hairons, $\omega_{0}$ is proportional to the energy gap:
	\begin{equation}
		\label{omemep}
		\omega_{0}\sim \epsilon \sim \frac{M_\textup{P}}{N^{3/2}}\sim \frac{T_\textup{dS}}{N} \ll 1\, . 
	\end{equation}
	The energy density stored in the condensate is
	\begin{equation}
		\rho_{\text{hairon}} \sim \Lambda M_\textup{P}^2,
	\end{equation}
	matching the de Sitter vacuum energy density, while, ignoring dynamical effects,
	the condensate has a pressure $p_{\text{hairon}} \sim - \Lambda M_\textup{P}^2$.
	
	\subsection{Holographic Scaling and Infrared Dynamics}
	
	Standard Quantum Field Theory treatments of the radiative corrections to the CC neglect the holographic structure of information in a dS spacetime. Rather, a consistent framework requires background fields supporting the horizon's quantum hairs, i.e., the hairon fields $\varphi_h(x)$. In this paradigm, the dS vacuum manifests as a coherent state
	\begin{equation}
		|\text{dS}\rangle = |\varphi_{h_1}, \dots, \varphi_{h_N}\rangle,
	\end{equation}
	with hairon properties including
	
	\begin{itemize}
		\item The formation of a weakly interacting Bose--Einstein condensate supporting coherent vacuum fluctuations;
		\item A characteristic energy gap $\epsilon \sim \sqrt{\Lambda}/N$ with collective energy $N\epsilon$ matching the CC~scale;
		\item Thermal background modifications to standard Quantum Field Theory bubble diagrams, enabling the stabilization of the CC~\cite{Addazi:2020axm,Addazi:2020wnc,Addazi:2020mnm}.
	\end{itemize}
	
	As demonstrated in the previous section, hairons emerge in a natural way as the moduli fields of orbifold gravitational instantons, providing a derivation of holographic degrees of freedom from first principles. The information-theoretic approach converts the CC problem from a fine-tuning puzzle to a question about the fundamental origin and dynamics of holographic information in the quantum gravitational vacuum.

	\section{Topological Foundation of Neutrino Condensation}
	\label{sec:topological_foundation}
	
	In this section we provide a short review, though with a different formalism, of the neutrino mass mechanism from gravitational anomalies, based on Refs.~\cite{Dvali:2016uhn,Funcke:2019grs,Dvali:2021uvk}. We will start by reviewing some aspects of QCD, which we will then extend to quantum gravity.
	
	\subsection{Chiral Symmetry and Axions in QCD}
	
	The topological susceptibility $\chi$ serves as the essential gauge for quantifying topological quantum fluctuations within the QCD vacuum~\cite{Addazi:2022whi}. Its formal definition is given by the variance of the global topological charge per unit volume,
	\begin{equation}
		\chi= \lim_{V \to \infty} \frac{\langle Q^2 \rangle}{V},
	\end{equation}
	where the integer-valued topological charge $Q$ in a four-volume $V$ is
	\begin{equation}
		\label{eq:Qt}
		Q=\frac{g^2}{32 \pi^2} \epsilon_{\mu\nu\lambda\sigma} \int d^4 x \ \mathrm{Tr}\left[ F^{\mu\nu}(x) F^{\lambda\sigma}(x) \right].
	\end{equation}
	Here, the field strength tensor $F_{\mu\nu} = T^a F_{\mu\nu}^a$ is matrix-valued, with the generators normalized such that $\mathrm{Tr} (T^a T^b) = \delta^{ab}/2$.
	
	In the low-temperature regime ($T < T_c \approx 150$ MeV), $\chi$ is intimately connected to the chiral condensate $\sigma$,
	\begin{equation}
		\sigma = -\lim_{m_q \to 0} \lim_{V \to \infty} \frac{1}{V} \int d^4 x \ \langle \bar{q} q(x) \rangle,
	\end{equation}
	which acts as the order parameter for spontaneous chiral symmetry breaking. The non-vanishing nature of $\sigma$ is responsible for generating the bulk of the mass for visible matter in the universe.
	
	Within leading-order chiral perturbation theory for two light flavors, a pivotal relationship emerges~\cite{Leutwyler:1992yt}:
	\begin{equation}
		\label{eq:LS}
		\chi = \frac{\sigma}{ m_u^{-1} + m_d^{-1} },
	\end{equation}
	where $m_u$ and $m_d$ are the masses of the up and down quarks,  respectively. This expression reveals a direct proportionality between the topological susceptibility and the chiral condensate. Consequently, significant topological fluctuations are a prerequisite for chiral symmetry breaking. In a scenario where $\chi$ vanished, $\sigma$ would also be zero, leaving chiral symmetry intact; this would result in nucleon masses on the order of 10 MeV, drastically lighter than their observed value of approximately 940 MeV. Furthermore, a nonzero $\chi$ explicitly breaks the $U(1)_A$ symmetry, offering a solution to the so-called $U(1)_A$ problem by explaining the anomalously large mass of the $\eta'$ meson compared to the octet of pseudo-Nambu--Goldstone bosons~\cite{tHooft:1976rip, Witten:1979vv, Veneziano:1979ec}.
	
	For temperatures below $T_c$ and small quark masses, chiral perturbation theory posits that the temperature dependence of $\chi_t(T)$ follows that of $\sigma(T)$, allowing for predictions of the former based on its zero-temperature value~\cite{Gasser:1986vb, Gasser:1987ah, Gerber:1988tt, Hansen:1990yg}.
	
	Above the critical temperature ($T > T_c$), chiral symmetry is restored ($\sigma(T)=0$). A subject of ongoing investigation is whether the $U(1)_A$ symmetry remains broken until a higher temperature $T_1 \gtrsim T_c$. A fascinating possibility is the existence of a temperature window $(T_c, T_1)$ where topological fluctuations ($\chi(T) \neq 0$) persist in the absence of chiral symmetry breaking ($\sigma(T)=0$). Elucidating the microscopic dynamics enabling this scenario is a key research objective.
	
	The temperature dependence of $\chi$ is also critically important for cosmology. It directly determines the properties of the axion, a well-motivated dark matter candidate. The axion field arises from the spontaneous breaking of a postulated global $U(1)_{\textup{PQ}}$ symmetry at a high-energy scale denoted by $f_A$~\cite{Peccei:1977hh, Weinberg:1977ma, Wilczek:1977pj}. The Peccei--Quinn mechanism not only dynamically resolves the strong CP problem, but also yields a cold dark matter candidate. The axion mass at a cosmological temperature $T$ is given by
	\begin{equation}
		m_A(T) = \frac{\sqrt{\chi(T)}}{f_A}.
	\end{equation}
	This mass is a fundamental parameter in the equations governing the axion field's evolution in the early Universe. The resulting relic abundance, produced primarily via the misalignment mechanism~\cite{Dine:1981rt, Preskill:1982cy, Abbott:1982af}, is therefore sensitive to the detailed thermal behavior of~$\chi(T)$.
	
	Computing $\chi(T)$ from first principles necessitates a non-perturbative approach. Lattice QCD simulations are the primary tool for this task, though the numerical challenge grows with temperature. At high $T$, topological fluctuations are exponentially suppressed, demanding immense statistical precision for reliable measurements. Current state-of-the-art direct lattice calculations extend to temperatures of around $550$ MeV. In the asymptotic high-temperature regime ($T \gg T_c$), the Dilute Instanton Gas Approximation (DIGA) becomes applicable, predicting a power-law decay $\chi(T) \propto T^{-(7 + N_f/3)}$ for $N_f$ quark flavors~\cite{Gross:1980br}.
	
	Ongoing lattice investigations relevant for axion cosmology employ various quark content ($N_f=0$, $2+1$, $2+1+1$) and discretization schemes, including staggered, Wilson, and twisted-mass fermions~\cite{Berkowitz:2015aua, Kitano:2015fla, Borsanyi:2015cka, Bonati:2015vqz, Petreczky:2016vrs, Borsanyi:2016ksw, Burger:2018fvb}. For comprehensive overviews of this active research field, see Refs.~\cite{Moore:2017ond, Lombardo:2020bvn}.

	\subsection{Topological Gravitational Sector}
	
	We now posit that the gravitational vacuum possesses a homologous topological structure. The analogous quantity to the QCD topological charge \eqref{eq:Qt} is precisely the gravitational instanton number, which is related to the integral of the Pontryagin density:
	\begin{equation}\label{QP}
		Q_G = p \int d^4 x \ R\tilde{R},\qquad R\tilde R\equiv \frac{1}{2}\epsilon^{\mu\nu\rho\sigma}R^\alpha_{\beta\mu\nu}R^\beta_{\alpha\rho\sigma},
	\end{equation}
	with $R^\alpha_{\beta\mu\nu}$ being the curvature tensor and $p$ a numerical normalization prefactor. This allows for the definition of a \textit{gravitational topological susceptibility}:
	\begin{equation}
		\label{eq:chi_grav}
		\chi_{G} = \lim_{V \to \infty} \frac{\langle Q_G^2 \rangle}{V}.
	\end{equation}
	A nonzero $\chi_G$ would signal robust topological fluctuations in the quantum gravity vacuum.
	
	\textls[-15]{In direct analogy to the QCD $\eta'$ meson, we introduce a gravitational pseudo-Nambu-- ~~~~} Goldstone boson $\eta'_G$. This field is the gravitational counterpart associated with the anomalous breaking of a chiral symmetry in the gravitational sector. Its mass is sourced by gravitational instantons, and the low-energy relation, mirroring the Witten--Veneziano mechanism, is
	\begin{equation}
		\label{eq:chiG_etaG}
		\chi_{G} = \frac{f_G^2}{N_f} m_{\eta'_G}^2 \, ,
	\end{equation}
	where $f_G$ is the decay constant of the $\eta'_G$ boson and $N_f$ is an effective number of degrees of freedom. This establishes $\chi_G$ as the gravitational order parameter analogously to the QCD topological susceptibility.
	
	The coupling to massless neutrinos dramatically alters this topological structure through the axial neutrino current anomaly. For a single neutrino species, the axial current
	\begin{equation}
		\label{eq:axial_current}
		j_5^\mu = \bar{\nu}\gamma^\mu \gamma_5\nu
	\end{equation}
	satisfies the anomalous divergence relation
	\begin{equation}
		\label{eq:anomaly}
		\partial_\mu j_5^\mu = p R\tilde{R} + 2im_\nu \bar{\nu}\gamma_5\nu\, ,
	\end{equation}
	where $m_\nu$ is the bare neutrino mass.
	
	The gravitational topological susceptibility $\chi_G$ and the neutrino condensate are related through a formula that mirrors the QCD relation \eqref{eq:LS}:
	\begin{equation}
		\label{eq:chiG_nu}
		\chi_{G} = \kappa m_{\nu}\langle \bar{\nu}\nu \rangle \, ,
	\end{equation}
	where $\kappa$ is a proportionality constant. This final relation completes the analogy: just as $\chi_{\text{QCD}}$ is tied to the quark condensate $\sigma$ and is essential for nucleon masses, so $\chi_G$ is tied to the neutrino condensate $\langle \bar{\nu}\nu \rangle$ and is essential for neutrino masses. The non-trivial topology of the gravitational vacuum thus provides the foundation for a dynamical mechanism generating neutrino masses.

	The neutrino condensate and its fluctuations can be parameterized using the composite field formalism:
	\begin{equation}
		\label{eq:composite_field}
		\mathcal{N} = \nu \bar{\nu} = \langle \nu \bar{\nu} \rangle e^{i\phi/f_\nu} = v e^{i\phi/f_\nu},
	\end{equation}
	where $\phi$ and $f_\nu$ are, respectively, the Nambu--Goldstone boson and the $\nu'_G$ decay constant. The effective Lagrangian then becomes
	\begin{equation}
		\label{eq:effective_interactions}
		\mathcal{L}_{\text{eff}} = g_v v  \bar{\nu}\nu + \sum_k \frac{\partial_\mu \phi_k}{f_\nu}  \bar{\nu} \gamma^\mu\gamma_5 \nu + g_{\eta_\nu} \eta_\nu  \bar{\nu} \gamma_5 \nu\, ,
	\end{equation}
	where $k$ \textls[-15]{is the sum over the Nambu--Goldstone boson species. The neutrino mass emerges as}
	\begin{equation}
		\label{eq:nu_mass_condensate}
		m_\nu = g_v v  \sim 1\, \text{meV}\, . 
	\end{equation}
	To be more precise, we can easily generalize this formalism to include all neutrino species. The interaction terms will read as follows:
	\begin{equation}
		\label{eq:enhanced_interactions}
		\mathcal{L}_{\text{int}} = \sum_{k=1}^{N_f^2-1} \frac{\partial_\mu \phi_k}{f_\nu} \sum_{i,j} g_{\phi,ij}  \bar{\nu}_i \gamma^\mu\gamma_5 \nu_j + \eta_\nu \sum_{i,j} g_{\eta_\nu,ij}  \bar{\nu}_i \gamma_5 \nu_j \,.
	\end{equation}

	\subsection{Gravitational Analog: Thermal Evolution and Neutrino Mass}
	
	In direct analogy to QCD, we propose that the gravitational topological susceptibility $\chi_G$ exhibits a characteristic thermal behavior. Below a critical temperature $T_G$ associated with gravitational symmetry breaking, $\chi_G(T)$ follows the behavior of the neutrino condensate $\langle \bar{\nu}\nu \rangle(T)$:
	\begin{equation}
		\chi_G(T) \propto \langle \bar{\nu}\nu \rangle(T) \quad \text{for} \quad T < T_G.
	\end{equation}
	Above $T_G$, the neutrino chiral symmetry is restored ($\langle \bar{\nu}\nu \rangle(T) = 0$). However, gravitational topological fluctuations may persist up to a higher temperature $T_2 \gtrsim T_G$, creating a window $(T_G, T_2)$ where
	\begin{equation}
		\chi_G(T) \neq 0 \quad \text{while} \quad \langle \bar{\nu}\nu \rangle(T) = 0.
	\end{equation}
	This scenario would imply that a non-trivial gravitational topology can exist independently of neutrino condensation at high temperatures, mirroring the potential $U(1)_A$ persistence found in QCD.
	
	The temperature dependence of $\chi_G(T)$ plays a crucial role in determining the neutrino mass throughout cosmic history. The dynamical neutrino mass at temperature $T$ is given~by
	\begin{equation}
		\label{eq:nu_mass_T}
		m_\nu(T) = \frac{\sqrt{\chi_G(T)}}{f_\nu},
	\end{equation}
	The thermal behavior of $m_\nu(T)$ represents a fundamental input for the evolution of the neutrino condensate field in the early Universe.
	
	Computing $\chi_G(T)$ from first principles is a formidable challenge in quantum gravity. Non-perturbative approaches such as lattice quantum gravity or holographic methods would be required, though these face even greater numerical challenges than in lattice QCD due to the exponential suppression of topological fluctuations at high temperatures.
	
	In the asymptotic high-temperature regime ($T \gg T_G$), a gravitational analog of the DIGA may become applicable. This would predict a power-law decay
	\begin{equation}
		\label{eq:chiG_highT}
		\chi_G(T) \propto T^{-\alpha},
	\end{equation}
	where the exponent $\alpha$ is determined by the properties of gravitational instantons. Ongoing investigations in various approaches to quantum gravity aim to determine this thermal behavior and its implications for neutrino cosmology.
	
	The resulting relic abundance of the neutrino condensate, produced through a misalignment-like mechanism in the early Universe, is therefore sensitive to the detailed thermal history of $\chi_G(T)$. This provides a direct link between gravitational topology, neutrino mass generation and potential cosmological implications.
	
	\vspace{0.1cm}
	
	Why are neutrinos uniquely sensitive to the hairon condensate and the gravitational topology? First, neutrinos are the only Standard Model fermions that are electrically neutral and lack strong charges, allowing unscreened couplings to the neutral hairon field. Second, their masslessness in the Standard Model makes them susceptible to the gravitational anomaly $\partial_{\mu} j_5^\mu \sim R\tilde{R}$, which ties their mass generation to the topological susceptibility $\chi_G$. Other fermions acquire masses at the electroweak scale via the Higgs mechanism, shielding them from the meV-scale dynamics of the hairon condensate. Thus, the small neutrino mass becomes a direct probe of the holographic vacuum structure.

	\section{Neutrino Condensation and Superfluidity}
	
	To follow up the discussion on the broad implications of our framework for neutrino Physics, we now consider a scenario in which both non-relativistic and relativistic neutrinos acquire an effective mass $m_{\mathrm{eff}} \simeq 1\,\mathrm{meV}$. However, only non-relativistic neutrinos with kinetic energies $E < m_{\mathrm{eff}} $ undergo condensation. Relativistic neutrinos, with energies $E \gg m_{\mathrm{eff}} $, remain largely unbound from the condensate and are predominantly observed in astrophysical contexts. The next section is then entirely dedicated to the phenomenological implications. It is important to distinguish the roles of hairons and neutrinos in the cosmological energy budget. The hairon condensate itself provides the dark energy density $\rho_{\text{DE}} \sim \Lambda M_{\text{P}}^{2}$. Neutrino condensation, mediated by couplings to hairons and to gravitational anomalies, forms a separate superfluid phase that may contribute to cold dark matter. The relation $m_{\nu} \sim \rho_{\text{DE}}^{1/4}$ emerges from the holographic scaling $m_{\nu} \sim M_{\text{P}}/N^{1/4}$ and does not imply that neutrinos are the source of dark energy; rather, it reflects a common origin in the topological structure of quantum gravity.
	
	\subsection{Non-Relativistic Effective Theory}
	
	In the non-relativistic regime we employ the formalism of fixed-axis spin states. Following the Refs.~\cite{Addazi:2016oob,Addazi:2022kjt}, we introduce the non-relativistic spin-up and spin-down wave functions $\psi_{\uparrow\downarrow}$, whose time derivatives are small compared to $m_{\mathrm{eff}}$. The kinetic term can be expressed in the Schr\"{o}dinger field form as
	\begin{equation}
		\sum_{s=\uparrow,\downarrow} \psi_{s}^{*} \left( i\partial_{t} + \frac{\nabla^{2}}{2m_{\mathrm{eff}}} \right) \psi_{s}(x),
	\end{equation}
	while the interaction term reduces to $G'_{F} \psi_{\uparrow}^{*} \psi_{\downarrow}^{*} \psi_{\downarrow} \psi_{\uparrow}$, where $G_{F}'\sim 1/m_\mathrm{eff}^2$ is an effective Fermi-like constant.
	
	Including a chemical potential $\mu$, the action for cold neutrinos thus reads as
	\begin{equation}
		\mathcal{S}[\psi^{\dagger},\psi] = \int d^{4}x \left[ \sum_{s=\uparrow,\downarrow} \psi_{s}^{*} \left( i\partial_{t} + \frac{\nabla^{2}}{2m_{\mathrm{eff}}} + \mu \right) \psi_{s}(x) + G'_{F} \psi_{\uparrow}^{*} \psi_{\downarrow}^{*} \psi_{\downarrow} \psi_{\uparrow}(x) \right], \label{eq:action-cooper}
	\end{equation}
	where ``$\dagger$'' denotes Hermitian conjugation. This structure resembles the BCS theory with an attractive interaction.
	
	\subsection{Hubbard--Stratonovich Transformation and Gap Equation}
	
	To study superfluidity, we introduce an auxiliary complex scalar field $\Phi(x)$ via the Hubbard--Stratonovich transformation. The partition function becomes
	\begin{eqnarray}
		Z & = & \frac{1}{Z_{0}} \int \mathcal{D}\psi \mathcal{D}\psi^{*} \mathcal{D}\Phi \mathcal{D}\Phi^{*} \, \exp\left( i\mathcal{S}'\left[\psi,\psi^{*},\Phi,\Phi^{*}\right] \right), \label{eq:functional-HS} \\
		\mathcal{S}'\left[\psi,\psi^{*},\Phi,\Phi^{*}\right] & = & \mathcal{S}_{0}\left[\psi,\psi^{*}\right] + \int d^{4}x \left[ -\frac{|\Phi|^{2}}{G'_{F}} - \Phi \left( \psi_{\uparrow}^{*} \psi_{\downarrow}^{*} \right) - \Phi^{*} \left( \psi_{\downarrow} \psi_{\uparrow} \right) \right], \label{eq:partition-HS}
	\end{eqnarray}
	where $Z_{0}$ is the partition function of the free theory. The equation of motion for $\Phi$ yields $\Phi = G'_{F} \psi_{\uparrow} \psi_{\downarrow}$, identifying $\Phi$ as the neutrino pairing field.
	
	Using the Nambu--Gorkov basis $\chi^{T} = \left( \psi_{\uparrow}, \psi_{\downarrow}^{*} \right)$, the action becomes
	\begin{equation}
		\mathcal{S}'\left[\chi,\chi^{\dagger},\Phi,\Phi^{*}\right] = \int d^{4}x \left( \chi^{\dagger} \mathcal{M}^{-1} \chi - \frac{|\Phi|^{2}}{G'_{F}} \right), \label{eq:action-HS}
	\end{equation}
	with an inverse propagator given explicitly by
	\begin{equation}
		\mathcal{M}^{-1} = \left( \begin{array}{cc}
			i\partial_{t} + \frac{\nabla^{2}}{2m_{\mathrm{eff}}} + \mu & -\Phi \\
			-\Phi^{*} & i\partial_{t} - \frac{\nabla^{2}}{2m_{\mathrm{eff}}} - \mu
		\end{array} \right). \label{eq:M-matrix}
	\end{equation}
	Integrating out the fermionic fields yields the one-loop effective action
	\begin{eqnarray}
		\frac{Z}{Z_{0}} & = & \int \mathcal{D}\Phi \mathcal{D}\Phi^{*} \, e^{i\mathcal{S}_{\mathrm{eff}}\left[\Phi,\Phi^{*}\right]}, \label{eq:eff-partition} \\
		\mathcal{S}_{\mathrm{eff}}\left[\Phi,\Phi^{*}\right] & = & -\frac{i}{2} \mathrm{Tr} \left[ \log \left( \mathcal{M}_{0} \mathcal{M}^{-1} \right) \right] - \int d^{4}x \frac{|\Phi|^{2}}{G'_{F}}. \label{eq:eff-action}
	\end{eqnarray}
	
	For a static, uniform ground state, the pairing field acquires a real vacuum expectation value $\Delta$. The saddlepoint approximation gives the gap equation
	\begin{equation}
		\Delta = \frac{G'_{F}}{4} \int \frac{d^{3}\mathbf{p}}{(2\pi)^{3}} \frac{\Delta}{\sqrt{ \left( \frac{\mathbf{p}^{2}}{2m_{\mathrm{eff}}} - \mu \right)^{2} + \Delta^{2} }}. \label{eq:gap-equation}
	\end{equation}
	A non-trivial solution $\Delta > 0$ always exists for $G'_{F} > 0$, indicating condensation. A BCS-like estimate then leads to
	\begin{equation}
		\Delta \sim p_{F} \exp\left( -\frac{1}{p_{F}^{2} G'_{F}} \right),
	\end{equation}
	where $p_{F} = \sqrt{2 \mu m_{\mathrm{eff}}}$ is the Fermi momentum. With $\mu \simeq m_{\mathrm{eff}} $, we find that $\Delta \sim m_{\mathrm{eff}}$. The critical temperature is also of the order of meV.
	
	\subsection{Effective Mass Generation and Symmetry Breaking}
	
	The condensate $\langle \Phi \rangle = \Delta$ implies a nonzero vacuum expectation value $\langle \xi \xi \rangle \simeq \Delta / G'_{F}$. This generates an effective Majorana mass term $\Delta (\xi^{\dagger} \xi^{\dagger} + \xi \xi) \simeq \Delta \, \nu^{T} \mathcal{C} \nu$, so that $m_{\mathrm{eff}} \simeq \Delta $.
	
	The superfluid state arises from spontaneous breaking of a global $U(1)$ symmetry in Equation\ \eqref{eq:action-cooper} in the ground state with $\Delta \in \mathbb{R}$. Fluctuations $\Phi(x) = (\Delta + \delta\rho(x)) e^{i\delta\theta(x)}$ yield a massive mode $\delta\rho$ and a massless Goldstone mode $\delta\theta$. Near the critical temperature, the dynamics are governed by the time-dependent Ginzburg--Landau Lagrangian
	\begin{equation}
		\mathcal{L}_{GL}[\Phi,\Phi^{*}] = i \Phi^{*} \partial_{t} \Phi - \frac{1}{2m_{\Phi}} \nabla \Phi^{*} \cdot \nabla \Phi - \alpha \Phi^{*} \Phi - \frac{1}{2} \beta (\Phi^{*} \Phi)^{2}, \label{eq:GL}
	\end{equation}
	where $m_{\Phi} \simeq 2 m_{\mathrm{eff}}$. Interestingly, the phase mode $\delta\theta$ exhibits a linear phonon-like dispersion relation between the angular frequency $\omega_{\mathbf{k}}$ and the wave number $\mathbf{k}$:
	\begin{equation}
		\omega_{\mathbf{k}}^{2} = -\frac{\alpha}{\beta} \frac{\mathbf{k}^{2}}{2m_{\Phi}} \sim \frac{\Delta}{m_{\mathrm{eff}}} \mathbf{k}^{2}.
	\end{equation}

	\section{Phenomenological Implications for Neutrino Physics}
	
	Neutrino superfluidity has rich phenomenological consequences, including the formation of neutrino vortices, exotic boson stars and distinctive gravitational wave signatures. The long-range interaction mediated by the massless mode could also modify the Newtonian potential, linking neutrino superfluids to dark matter phenomena~\cite{khoury2016dark, Addazi:2018ivg, Sharma:2018ydn, Ferreira:2018wup, Berezhiani:2018oxf, Famaey:2019baq}. These possibilities have been explored extensively in the literature~\cite{berezhiani2015theory, volovik2001superfluid, khoury2016dark, Addazi:2018ivg, Sharma:2018ydn, Ferreira:2018wup, Berezhiani:2018oxf, Famaey:2019baq,Addazi:2016oob,Addazi:2022kjt}, but here they are surprisingly shown to be related to the unified framework under examination.
	
	\subsection{Novel Cosmological Evolution and Weakened Mass Bounds}
	\label{subsec:cosmo_evolution}
	
	A time-varying neutrino mass, generated via a supercooled phase transition, dramatically alters the standard cosmological narrative. In this scenario, neutrinos remain effectively massless throughout most of the Universe's history, only acquiring their mass in the very recent cosmological past. This decouples them from early structure formation processes, leading to a significant relaxation of the cosmological bound on the sum of neutrino masses---potentially as high as $\sum m_\nu < 4.8$~eV. This revised upper limit provides a new target for next-generation laboratory experiments such as KATRIN and PTOLEMY~\cite{Dvali:2016uhn,Funcke:2019grs}. The model naturally induces a strong correlation between the neutrino and the dark energy sectors, as the latent energy of the false vacuum---which drives the cosmic \mbox{acceleration---converts} directly into neutrino masses at the epoch of the phase transition~\cite{Dvali:2016uhn}. However, later we will further comment on this possibility, as such considerations may be revisited in light of the $\mathcal{HN}$ framework.
	
	\subsection{A Topologically Complex Relic Neutrino Background}
	\label{subsec:topological_background}
	
	The postulated phase transition spontaneously breaks the neutrino flavor symmetry, leading to the formation of unique cosmological topological defects: global skyrmions, monopoles, strings, domain walls~\cite{Dvali:2021uvk}. These defects are exceptionally ``soft'' due to their low formation energy and are composed purely of Standard Model neutrino fields . Critically, their properties differ fundamentally depending on whether neutrinos are Dirac or Majorana particles, offering a potential indirect path to resolving this central question. Furthermore, the enhanced neutrino--neutrino interactions can lead to a strongly coupled cosmic neutrino background (C$\nu$B), which may form a superfluid, annihilate, or bind up. This could facilitate significant clustering within our galaxy, substantially enhancing the prospects for its direct detection in tritium beta-decay experiments 
	\cite{Dvali:2021uvk}.
	
	\subsection{Distinctive Astrophysical and Laboratory Signatures}
	\label{subsec:astro_lab_signatures}
	
	The framework predicts a suite of distinctive signatures across various observational frontiers.
	
	\begin{itemize}
		\item \textbf{Enhanced Neutrino Decays}: \textls[-35]{The model predicts fast neutrino decays (e.g., $\nu_i \to \nu_j + \phi$)} into $\phi$ particles, which are light Nambu--Goldstone bosons. This can cause measurable deviations from the expected equal-flavor ratio of high-energy astrophysical neutrinos observed at detectors like IceCube~\cite{Funcke:2019grs}. The decay kinematics and the resulting neutrino/antineutrino spectra provide a unique channel to distinguish between the Dirac or Majorana nature of neutrinos. The compatibility of this process with cosmological limits coming from the Big Bang nucleosynthesis (BBN) and the cosmic microwave background (CMB) was also discussed in great detail in Ref.~\cite{Funcke:2019grs}. Let us remark that the enhanced neutrino decays $\nu_i \to \nu_j + \phi$ must respect cosmological constraints from BBN and CMB. The decay rate scales as $\Gamma \sim m_{\nu}^3 / f_{\nu}^2$, where $f_{\nu}$ is the decay constant. For $m_{\nu} \sim \text{meV}$ and $f_{\nu} \gtrsim 10^8\ \text{GeV}$, the lifetime exceeds the age of the universe, ensuring no significant energy injection during BBN or CMB epochs. Such a large $f_{\nu}$ is natural in our framework, as it is tied to the Planck scale and the holographic suppression $1/N$. Thus, the model can satisfy cosmological bounds while producing observable flavor-ratio anomalies in high-energy neutrino telescopes.
		
		\item \textbf{New Short-range Forces}: The resulting neutrino condensate can provide a mass for a hypothetical $B-L$ gauge boson, giving rise to a new, gravity-competing force. This force would be detectable in high-precision, short-distance laboratory measurements~\cite{Dvali:2016uhn}. 
		
		\item \textbf{Dynamical Neutrino Mass}: The time- and space-dependent nature of the neutrino mass matrix, as a direct consequence of the topological defects network, could leave an imprint observable in future long-baseline neutrino oscillation experiments. Furthermore, the violent cosmological phase transition itself could source a stochastic gravitational wave background, providing a direct probe of the mass generation mechanism with current and future gravitational wave observatories~\cite{Dvali:2021uvk}.
		
		\item \textbf{Probing the Mass Mechanism}: In light of future neutrino experiments such as JUNO~\cite{An:2015jdp,Li:2014qca,Djurcic:2015vqa}, let us comment on the implications of a gravitational alternative to our model. JUNO is a middle-baseline antineutrino reactor experiment, which detects antineutrinos generated by nuclear power sources. Such a measurement will determine the neutrino mass hierarchy with a promised significance of $4\sigma$ after six years of data-taking. In particular, the high-resolution measurement of the antineutrino energy spectrum will enable a precise determination of the neutrino oscillation parameters, $\Delta m_{21}^{2}$, $\Delta m_{ee}^{2}$ and $\sin^{2} \theta_{12}$, with a precision of approximately $1\%$. This information is crucial for determining the sign of $\Delta m_{31}^{2}$, i.e., whether $m_3 > m_1$ (normal hierarchy) or $m_3 < m_1$ (inverted hierarchy).
		In a gravitational framework, the neutrino mass hierarchy is not set by particle couplings, but emerges from the interaction of overlapping neutrino wave functions with the hairons. This model can naturally accommodate both mass orderings. From this perspective, JUNO would be measuring a gravitational imprint on the neutrino mass spectrum, potentially influenced by the cosmic gravitational potential or dark energy. The mass hierarchy is linked to the geometry of spacetime, such that the ratio of mass eigenvalues is tied to a curvature parameter $\kappa$, i.e.,
		\begin{equation}\label{temp}
			\frac{m_3}{m_1} \propto \kappa_{31},
		\end{equation}
		where $\kappa_{31}$ characterizes the differential hairon effect on the respective mass eigenstates. Thus, an inverted hierarchy corresponds to a specific configuration $\kappa_{31} < 1$, while a normal hierarchy corresponds to $\kappa_{31} > 1$. The effective neutrino mass $m_{\nu}(T)$ in Equation\ \eqref{temp} depends on the gravitational topological susceptibility $\chi_G(T)$, which is a scalar background field, not the local spacetime curvature. This is analogous to the temperature-dependent quark condensate in QCD and does not violate the equivalence principle, because it does not introduce species-dependent couplings to the gravitational metric. The mass variation is adiabatic and uniform across all neutrino flavors, preserving the universality of the gravitational coupling.
	\end{itemize}
	
	Yet, the most impressive prediction may be the following. In the presence of a strong magnetic field, a photon could theoretically decay into a pair of neutrinos---a seemingly bizarre phenomenon~\cite{Addazi:2021ivu}. This idea connects the Peccei--Quinn axion to a specific particle state known as the ``domestic axion''~\cite{Dvali:2016eay,Addazi:2021ivu}. If valid, this hypothesis would have profound cosmological implications. For example, ultra-high-energy neutrinos observed in space could be produced when photons convert into these axions within cosmic magnetic fields, with the axions then decaying into neutrinos. Consequently, terrestrial experiments that search for photon-axion conversions in magnetic fields could find a direct correlation with signals in neutrino detectors.

	\section{Open Problems for Neutrino Mass from Gravitational Anomalies}
	
	While promising, the mechanism from Ref.~\cite{Dvali:2016uhn} faces four major challenges that limit its viability.
	
	\begin{description}
		\item[An Unexplained Scale] The theory does not account for the reason why the anomalous term $\langle R\tilde{R} \rangle$ must be as small as the observed vacuum energy; its value remains a free parameter.
		
		\item[Transferred Fine-tuning] The naturalness problem of the cosmological constant is not solved, but is instead passed on to the problem of neutrino mass. This makes it less competitive than other neutrino mass models (e.g., the see-saw mechanism) which are better understood and more stable.
		
		\item[A Prerequisite for Neutrino Mass] The mechanism only functions if neutrinos already have a small, bare mass, which itself requires an explanation for its unnatural smallness.
		
		\item[No Genuine Unification] A central flaw in this gravitational anomaly proposal is its failure to genuinely unify dark energy and neutrino mass. The mechanism requires the anomalous neutrino mass to be generated via a phase transition in the late Universe, coinciding with an average temperature of around the meV scale. However, this late-time activation entails that the proposed $R\tilde{R}$ term remains inactive during the CMB epoch. Since dark energy must already be influencing the Universe's expansion at that time, the model cannot explain its origin.
		
	\end{description}
	
	In the following section, we will show how combining this mechanism with the $\mathcal{HN}$ framework and gravitational instantons can eventually provide solutions to these open problems, introducing new phenomenological possibilities as well.

	\section{$\mathcal{HN}$ and Neutrino Mass}
	\label{sec:hn_neutrino_mass}
	
	The $\mathcal{HN}$ paradigm provides a unified framework that naturally explains the simultaneous stabilization of the CC and the origin of neutrino mass. Furthermore, it suggests an information-theoretic interpretation for the neutrino mass itself.
	
	\subsection{Neutrino Mass from the Holographic Principle}
	
	From an infrared cutoff on the volume of the Universe in a four-dimensional dS spacetime, the following relation is derived in the large-volume limit:
	\begin{equation}
		\label{eq:cutoff}
		\langle Q_{G} \rangle \simeq \chi_{G} V_{4} \simeq \langle R\tilde{R} \rangle \Lambda^{-2},
	\end{equation}
	where $Q_G$ denotes the gravitational topological charge and $V_4$ the dS spacetime volume, while the quantity $R\tilde{R}$ was defined in Equation\ \eqref{QP} in terms of the curvature tensor. Assuming this topological charge scales holographically with the number of degrees of freedom, i.e.,\
	\begin{equation}
		\label{eq:sccc}
		\langle Q_{G} \rangle \sim N,
	\end{equation}
	one arrives at a key result:
	\begin{equation}
		\label{eq:resultN}
		\langle R\tilde{R} \rangle \sim N \Lambda^{2} \sim M_{\textup{P}}^{2} \Lambda \sim \rho_{\text{vac}}.
	\end{equation}
	This directly links the vacuum expectation value of the Chern--Pontryagin density to the observed scale of dark energy, $\rho_{\text{vac}}$.
	
	This implies that the gravitational topological charge must be large, scaling holographically with the de Sitter entropy:
	\begin{equation}
		\label{eq:scaling}
		\langle Q_{G} \rangle \sim N \sim S_\textup{dS} \sim M_{\textup{P}}^{2} / \Lambda.
	\end{equation}
	Consequently, the neutrino mass scales as
	\begin{equation}
		m_{\nu} \sim \rho_\text{vac}^{1/4}\sim (M_\textup{P}^{2}\Lambda)^{1/4} \sim M_\textup{P}/N^{1/4}\, . 
	\end{equation}
	The smallness of the neutrino mass is therefore a direct consequence of the large topological charge, requiring indeed $N \sim 10^{120}$.
	
	\subsection{Topological Complexity in Vacuo}
	\label{subsec:topological_complexity}
	
	A simple, smooth de Sitter space cannot support such a large topological charge, as its entropy is large but its topological complexity is not. This indicates that the true infrared structure of quantum gravity is far more complex than that of a standard dS spacetime. Our main proposal---that the de Sitter entropy can be reconstructed from a large number of Euclidean conic singularities (instantons)---resolves this issue. Each singularity introduces a modulus, or ``hairon'', effectively increasing the genus of the spacetime manifold. A large-$N$ orbifold with a $\mathbb{Z}_N$ quotient can model this configuration, where the abundance of these topological defects provides the necessary topological complexity for matching the dS entropy. In particular, the key is thinking about the correspondence between conic singularities and Euclidean topological strings. Such strings in the dS bulk puncture the dS horizon, leading to the following result: 
	\begin{equation}
		\label{punnncc}
		\int_{S_{4}/\mathbb{Z}_{N}} d^{4}x R\tilde{R}=\int_{\Sigma}C_{G} = N\, ,
	\end{equation}
	where $\Sigma$ is the boundary surface. Thus, in a highly non-trivial way, the Euclidean action, the dS entropy and the topological term all scale as $N$, which in general is not true for all instantons except for only a mini-selected space of them. This insight seems to fundamentally relate the entropy of the dS spacetime with topological complexity.
	
	Such a portrait can suggest to conjecture a correspondence of the BH and the dS thermodynamic entropy with the topological entropy associated with braids and Wilson lines in the bulk:
	$$ {\it Holographic}\, {\it Entropy} =  {\it Topological}\, {\it Entropy}$$ 
	
	or, in short,
	$${\bf HE}={\bf TE}\, .$$ 
	
	The topological entropy~\cite{TE1,TE2} was studied for several reasons, including investigation of two-dimensional fluids~\cite{TE3}. It is also interesting to note that, typically, the topological entropy of the flow has an exponential growth rate of material lines~\cite{TE3}. We are tempted to conjecture that such exponentialization is on the number of punctures on the BH or dS horizon. Moreover, this could also be related to information scrambling and chaotization inside BHs. Indeed, the topological entropy could scale as the entanglement entropy, which in turn plays a prominent role in our understanding of BHs under the AdS/CFT correspondence. In fact, in our previous work~\cite{Addazi:2020mnm}, we stressed the possibility of relating BH and dS entropy to the topological entanglement entropy, which is in general not necessarily equivalent to the topological entropy. At the moment, we do not have enough computational power to test such a conjecture. In the future, computational methods for estimating the topological entropy, such as the one proposed in~\cite{TEA,TEB,TEC}, could help to sustain the ${\bf HE}={\bf TE}$ conjecture.
	
	For a better understanding of the conjecture, let us list a series of mathematical facts about braids and topological entropy~\cite{A1}: (i) periodic braids have topological entropy equal to zero; (ii) reducible braids that can be reduced into components which are all periodic, have a topological entropy equal to zero; (iii) all other braids have a topological entropy which is positive and nonzero, and are dubbed pseudo-Anosov braids (PABs). Therefore, our claim is that gravitational Wilson lines around topological string defects, related to PABs, contribute to the BH entropy. Moreover, if 
	\begin{equation}
		\label{ST}
		S_{T}(\beta^{N})\leq N S_{T}(\beta)\sim N,
	\end{equation}
	where $S_{T}$ is the topological entropy, then $\beta$ is an irreducible PAB; this is exactly the desideratum for the BH entropy, whereas typically one finds that $S_{T}(\beta)\sim O(1)$. Thus, a hairon would correspond to a PAB or, in other words, hairons would correspond to topological entropy units.

	\subsection{Origin of Neutrino Mass and a Hairogenic Phase Transition}
	\label{subsec:neutrino_mass_origin}
	
	A second challenge is generating the neutrino mass itself. The mechanism requires that neutrinos have no significant bare mass. This is resolved by postulating a coupling between hairons (\(\varphi_h\)) and neutrinos (\(\nu\)):
	\begin{equation}
		\label{eq:coupling}
		\mathcal{L} \supset \varphi_h \nu \bar{\nu} + \text{h.c.},
	\end{equation}
	with a hairon vacuum expectation value \( \langle \varphi_h \rangle =  \rho_{\text{vac}}^{1/4} \). Since hairons constitute the dark energy background, this interaction naturally generates a meV-scale mass for neutrinos. Gravitational anomalies from $R\tilde R$ terms are seamlessly integrated into this picture and may yield further phenomenological signatures. The issue of radiative corrections to the neutrino mass is intrinsically tied to the stabilization of the CC, which is already addressed within the $\mathcal{HN}$ framework. To summarize, the neutrino mass hierarchies receive contributions from both gravitational anomaly terms $R\tilde{R}$ and the direct hairon--neutrino coupling of Equation\ \eqref{eq:coupling}. The neutrino bare masses generated from the hairon--neutrino coupling \eqref{eq:coupling} are dressed by contributions from gravitational anomalies, which would otherwise be ineffective. We assumed that the dominant contributions to the neutrino mass hierarchies come from gravitational anomalies, in a similar fashion to the considerations of Ref.~\cite{Funcke:2019grs}.
	
	\subsection{A Remark on Early-Universe Neutrino Superfluidity}
	\label{subsec:early_superfluidity}
	
	A remark on neutrino superfluidity is also in order. While the standard cosmological model suggests that it can only form in the late Universe, the $\mathcal{HN}$ picture offers an alternative scenario. If a significant population of neutrinos were produced non-relativistically in the very early Universe, their thermalization with the Standard Model plasma would be highly suppressed due to the weakness of their interactions (\(G_F E^2 \sim 10^{-14}\)).
	
	In the $\mathcal{HN}$ model, instead, since the neutrino mass is generated by interactions with the hairon dark energy field---which has a temperature \(T_\textup{dS}\) distinct from that of ordinary matter---a hairon-triggered neutrino phase transition could also occur in the early Universe. This framework thus provides a genuine unification of dark energy and neutrino mass, with a common origin potentially rooted in the early Universe's dynamics.

	\subsection{Neutrino Decoherence, Unitarity and the Hairon Sector}
	
	The $\mathcal{HN}$ framework suggests that neutrinos interact with a large number of hidden fields, the so-called ``hairons''. As hairons are not directly observable, their presence would be detected indirectly through decoherence effects in the neutrino sector. This provides a phenomenological signature that can be searched for as an apparent violation of unitarity in neutrino oscillations, a topic that was already extensively discussed in the literature (see, e.g., the Refs.~\cite{Addazi:2021xuf,AlvesBatista:2023wqm} for recent reviews).
	
	However, a crucial distinction must be made: even if the effective non-unitary formalism---based on the evolution of the neutrino density matrix rather than the wave function---is substantially the same, the underlying physics is fundamentally different. In the $\mathcal{HN}$ picture, unitarity and CPT symmetry are not fundamentally violated, as this framework is intrinsically linked to ideas resolving the BH and dS information paradoxes.
	
	Another important remark concerns the coupling strength. Contrary to other Standard Model particles, for which decoherence effects from quantum gravity are expected to be suppressed by the gravitational coupling $\alpha = G E^{2}$, Equation\ \eqref{eq:coupling} provides a \emph{direct} coupling between neutrinos and hairons. Despite this direct coupling, the relative fluctuations in the hairon field are highly suppressed, scaling as in a coherent state:
	\begin{equation}
		\label{scall}
		\frac{\Delta N }{\langle N\rangle} = \frac{1}{\sqrt{N}} \sim \frac{\sqrt{\Lambda}}{M_\textup{P}}\, .
	\end{equation}
	This renders any potential effect extremely small and challenging to detect.
	
	In the context of astrophysics and cosmology, ultra-high-energy neutrinos propagating over cosmological distances could experience cumulative effects from these fluctuations, potentially altering their propagation and inducing observable time delays. Nevertheless, for such effects to become of order $1$, the required propagation length would likely have to be comparable to the Hubble length. The search for potential amplifiers of these phenomena is an active subject of investigation, but beyond the purposes of this work.

	\section{Conclusions and Remarks}
	\label{sec:conclusion}
	
	In this work, we have put forth a unified framework which addresses two of the most pressing naturalness problems in fundamental Physics---the smallness of the cosmological constant and the origin of neutrino mass---by rooting them in the holographic and topological structure of quantum gravity.
	
	Our central thesis asserts that the information content of the de Sitter space, quantified by its entropy $S_{\textup{dS}} = N \sim M_{\textup{P}}^2/\Lambda$, is physically encoded in a vast number of light, coherent degrees of freedom: the ``hairons''. We have provided a first-principles derivation of these hairons by constructing a new class of gravitational instantons, the orbifolds $S^4/\mathbb{Z}_N$. The dimension of the moduli space of these instantons scales as $N$, hence we have identified these moduli with the proposed hairon fields. A crucial element is the emergence of a $\mathbb{Z}_N$ symmetry from the instanton background, which ensures the physical distinguishability of the $N$ hairons, leading to the correct entropy count without factorial over-counting. The dS vacuum is thus understood not as an empty space, but as a macroscopic coherent state---a Bose--Einstein condensate---of these fundamental holographic excitations.
	
	This holographic foundation provides the bedrock for a novel neutrino mass generation mechanism. The gravitational topological susceptibility $\chi_G$, tied to the neutrino condensate $\langle \bar{\nu}\nu \rangle$ via a relation analogous to the QCD chiral anomaly, scales holographically with the number of degrees of freedom: $\langle R\tilde{R} \rangle \sim N \Lambda^2$. This directly links the neutrino mass $m_\nu$ to the cosmological constant via an \textit{information see-saw mechanism}, as $m_\nu \sim M_\textup{P}/N^{1/4}$, naturally explaining the meV scale which is common to both. The required large topological charge is naturally supported by the complex topology of the $S^4/\mathbb{Z}_N$ orbifold instantons, where conic singularities provide the necessary topological complexity.
	
	This unified picture yields a rich and testable phenomenology, as summarized next.
	\begin{itemize}
		\item \textbf{Neutrino Superfluidity}: The framework predicts that non-relativistic neutrinos can form a superfluid condensate at meV energies, with potential implications for dark matter.
		\item \textbf{Topological Defects}: A phase transition in the neutrino sector could produce a spectrum of unique, soft topological defects (skyrmions, strings, domain walls) in the cosmic neutrino background.
		\item \textbf{Exotic Signatures}: The model predicts enhanced neutrino decays, new short-range forces, time-varying neutrino masses and potentially observable effects like photon-decay to two neutrinos in strong magnetic fields.
		\item \textbf{Apparent Unitarity Violation}: The direct coupling of neutrinos to the hairon environment induces decoherence, which would be detected as an apparent violation of unitarity in neutrino oscillations, a distinctive signature of the holographic environment.
	\end{itemize}

	Furthermore, our framework suggests several profound implications and avenues for future research, which are listed next.
	
	\begin{itemize}
		\item \textbf{De Sitter Stability and Swampland}: The $\mathcal{HN}$ picture provides a microscopic rationale for the apparent metastability of the de Sitter space. The exponential suppression of vacuum decay, $\Gamma \sim e^{-N}$, is a natural consequence of the vast information content, potentially offering a new perspective to reconcile the swampland conjectures with observations. Moreover, swampland bounds may be related to the holographic information scrambling time. 
		\item \textbf{Dynamical Dark Energy}: The coherent state description naturally accommodates a slowly evolving hairon condensate, suggesting that the equation of state $w(t)$ of dark energy could be dynamically determined by the underlying hairon dynamics.
		\item \textbf{The HE=TE Conjecture}: We have proposed a deep connection between holographic entropy (HE) and topological entropy (TE), suggesting that the BH and dS entropy may be counted by the topological complexity of braids and Wilson lines in the gravitational bulk. Investigating this conjecture represents a major future endeavor.
	\end{itemize}
	
	In conclusion, we have presented a concrete framework wherein the cosmological constant and the neutrino mass are not independent, fine-tuned parameters but rather emergent, intertwined consequences of the holographic information structure of spacetime. The CC problem is thus transmuted into the question of why the Universe possesses $N \sim 10^{120}$ holographic degrees of freedom, while the neutrino mass scale is elegantly set by an information-based see-saw mechanism. This work opens up new pathways toward understanding the quantum nature of the cosmic acceleration as well as the generation of matter, unifying the infrared and ultraviolet puzzles of contemporary Physics through the lens of quantum gravity.

	\vspace{0.2cm}
	
	{\bf Acknowledgements}.
	During the preparation of this work, the authors used DeepSeekV3 to improve readability and language. After using this tool, the authors reviewed and edited the content as needed and take full responsibility for the content of the publication. AA's work is supported by the National Science Foundation of China (NSFC) through the grant No.\ 12350410358; the Talent Scientific Research Program of College of Physics, Sichuan University, Grant No.\ 1082204112427; the Fostering Program in Disciplines Possessing Novel Features for Natural Science of Sichuan University, Grant No.\ 2020SCUNL209 and 1000 Talent program of Sichuan province 2021. GM acknowledges the support of Istituto Nazionale di Fisica Nucleare (INFN), Sezione di Napoli, Iniziativa Specifica QGSKY. This paper is based upon work from COST Action CA21136---Addressing observational tensions in cosmology with systematics and fundamental physics (CosmoVerse), supported by COST (European Cooperation in Science and Technology).

\end{document}